\documentstyle[preprint,prb,aps]{revtex}
\tightenlines
\begin{document}
\draft
\title{Far-infrared spectroscopy of nanoscopic InAs rings}
\author{ Agust\'{\i} Emperador, Mart\'{\i} Pi and
Manuel Barranco}
\address{Departament ECM,
Facultat de F\'{\i}sica,
Universitat de Barcelona, E-08028 Barcelona, Spain}
\author{ Axel Lorke}
\address{Sektion Physik, LMU M\"unchen -
Geschwister-Scholl-Platz 1
80539 M\"unchen, Germany}
\date{\today}

\maketitle

\begin{abstract}

We have employed  time-dependent local-spin density theory
to analyze the far-infrared transmission spectrum of  InAs
self-assembled nano-rings recently reported [A. Lorke
et al, cond-mat/9908263 (1999)]. The overall agreement
between theory and experiment is good, which on the one hand
confirms that the experimental peaks indeed reflect the ring-like
structure of the sample, and on the other hand, asseses the suitability
of the theoretical method to describe such small nanostructures.
The addition energies of one- and two-electron rings are also
reported and compared with the corresponding capacitance spectra.

\end{abstract}
\narrowtext
\section*{}

Recent progress in nanofabrication techniques\cite{Lor98,Lor99}
has allowed to construct
self-assembled nanoscopic InGaAs quantum rings occupied with one
or two electrons each, and submitted to perpendicular magnetic
fields $(B)$ of up to 12 T. 
These are the first spectroscopic data available on rings
in the scatter-free, few electrons limit in which quantum effects
are best manifested. Previous spectroscopic studies dealt with
microscopic rings\cite{Dah93} in GaAs-Ga$_x$Al$_{1-x}$As
heterostructures, fairly well reproduced by classical or
hydrodynamical models\cite{Pro92,Zar96}.

In spite of the lacking of experimental information, the
study of nanoring structures has already attracted  a
strong theoretical interest\cite{Cha94,Gud94,Wen96,Hal96,Tan99}. 
We recall
that due to the non-applicability of the generalized Kohn theorem, a
very rich spectroscopic structure is expected to appear in
few electrons nanorings,
as anticipated by Halonen et al\cite{Hal96} and
also found in recent works\cite{Emp99,Mag99}.

In this paper we attempt a quantitative description of some 
spectroscopic
and ground state (gs) properties of the experimentally studied
nanorings\cite{Lor98,Lor99} using current-density  (CDFT) and
time-dependent local-spin density (TDLSDFT) functional theories.
The reason for such an attempt is twofold.
On the one hand, to contribute to put on a firmer
basis the interpretation of current experiments as manifestation
of actual properties of few-electrons ring-shaped nanostructures.
On the other hand, to disclose the capabilities and
limitations of density functional methods to describe such small
systems.

Following Ref. \onlinecite{Cha94}, we have modeled the
ring confining potential by a parabola
\begin{equation}
V^+(r)=\frac{1}{2}\, m \,\omega_0^2 \,(r - R_0)^2
\label{eq1}
\end{equation}
with\cite{Lor99} $R_0$= 14 nm and the frequency  $\omega_0$
fixed to reproduce the high energy  peak found in the 
far-infrared (FIR) transmission  spectrum
at $B$= 0. For $N$= 2 electrons this yields
$\omega_0\sim$ 12.3 meV. The electron effective mass
$m^*$= 0.063 (we write $m = m^* m_e$ with $m_e$ being the
physical electron
mass) and effective gyromagnetic factor $g^*= -0.43$
 have been taken from the experiments\cite{Fri96,Lor96,Mil97},
 and the value  of the dielectric constant has been 
 taken to be $\epsilon$= 12.4.

To obtain the structure of the gs we have resorted to CDFT as
described in Refs. \onlinecite{Fer94,Pi98}, and to obtain the
charge density response we have used TDLSDFT as described in
Ref. \onlinecite{Ser99}, which has been recently applied
to the ring geometry \cite{Emp99}.
It is worthwhile to point out that we have not found
any significant difference between using CDFT or LSDFT to describe the
gs of the studied rings in the range of $B$ values of the
present work. The suitability of CDFT to describe such a small
electronic system has been shown by Ferconi and Vignale\cite{Fer94}
comparing the results obtained for a dot with $N$= 2 electrons with
exact and Hartree-Fock calculations. We refer the  reader
to the mentioned references for a detailed exposure of the methods.

The results obtained for the $N$= 2 ring are presented in Figs.
\ref{fig1}-\ref{fig4}. We have used a small temperature $T$= 0.1 K
to work them out. Figure \ref{fig1} shows that the ring becomes
polarized near $B$= 3 T. Besides, two other $B$-induced changes
arise in the gs at $B\sim$ 8 T and, more weakly, at $B\sim$ 14 T. These
changes can be traced back to sp level crossings
\cite{Lor99}. As displayed in Fig. \ref{fig2}, the changes in the
$B$-slope appear when an occupied sp level is substituted by
an empty one. At  $B\sim$ 8 T, this involves the substitution of
the $l$= 0 sp level by the $l$= 2 one, and at $B\sim$ 14 T
the $l$= 1 sp level is substituted by the $l$= 3 one\cite{note}.
Other level crossings do not involve such substitutions, but a different
ordering of the occupied levels and do not seem to produce  a
substantial effect (see for instance the crossings at $B\sim$ 6 and
$\sim$ 11.5 T).

The experimentally observed change in the FIR spectrum 
around $B$= 8 T has been attributed\cite{Lor99}
to the crossing of $l$= 0 and 1 sp levels on the basis of a simple
single-electron model (see also Fig. \ref{fig5}).
A realistic description of the crossings requires
to incorporate in the theoretical description the spin degree of freedom,
of which single electron or Hartree models\cite{Mag99} lack
whereas CDFT or LSDFT do not. Yet, we
confirm the finding\cite{Lor99} that a magnetic induced transition
takes place in the gs when approximately 1 flux quantum penetrates
the effective interior area of the ring at $B\sim$ 8 T, and predict
another one at $B\sim$ 14 T when this area is penetrated by $\sim$
2 flux quanta.

The changes in $B$-slope of the total energy correlate well with
these in the electronic chemical potential (the energy of the
last occupied sp level in Fig. \ref{fig2}). The gross structure
of the chemical potential and total energy displays the
well known periodic, Aharonov-Bohm-type oscillation found in extreme
sp models\cite{Cha94,Lor99}:

\begin{equation}
\epsilon_l =\frac{\hbar^2}{2 \,m \,R^2_0}
\left( l - \frac{e}{\hbar c} \,R^2_0 \,B \right)^2 \,\,\, .
\label{eq2}
\end{equation}

The experimental FIR resonances have been grouped\cite{Lor99}
into different modes using a different symbol for each
group. Here, we have used the same symbol to represent the experimental
resonances in Figs. \ref{fig3} and \ref{fig4}.
Figure \ref{fig3} shows the dipole charge density strength
function in an arbitrary logarithmic scale as a function of the 
excitation energy. The curves have been offset for clarity.
Charge density excitations (CDE) can be identified as 
`ridges' in the plot, allowing to make a sensible comparison
with experiment  not only of the peak energies themselves, but 
also of the way the experimental modes have been grouped.
 A plot of the more intense CDE's is presented in Fig. \ref{fig4}
 as a  function of $B$, which is
qualitatively similar to that of Halonen et al\cite{Hal96} for
an $N$= 2 quantum dot with a repulsive gaussian impurity in its
center. For completeness, we also show in Fig. \ref{fig3} the 
longitudinal spin density strength function\cite{Ser99} for the cases
in which the ring is not fully polarized. As both strengths
coincide in the non-interacting case, the observed shifts are a
measure of
the importance of the electron-electron interaction, which affects
more the low energy peaks than the high energy ones.

These figures  show that the FIR dipole
response is splitted
into two large groups of peaks. The low energy peaks correspond
to transitions involving only $n$= 0 sp levels and are
 $\Delta n$= 0 transitions, whereas the high energy peaks
involve $n$=0 and 1 sp levels and are $\Delta n$= 1 transitions.
One can  easily distinguish
two sets of resonances, a low-lying $\Delta n$= 0 one,
and a high-lying $\Delta n$= 1  one
exhibiting the usual Zeeman splitting when a magnetic field is
applied. The intensity of the high energy resonance is 
more than one order of magnitude smaller than that of the low 
energy one.
Experimentally, both sets have similar oscillatory strengths,
whereas TDLSDFT yields a $\sim$ 90-10 \% share at most. The
calculations in Ref. \onlinecite{Hal96} also yield rather
different absorption intensities to these resonances.
We have checked that the computed spectrum fulfills the $f$-sum 
rule\cite{Ser99} to within $\sim 98 \%$, thus leaving no room for
higher energy, $\Delta n >$ 1 peaks to appear within TDLSDFT.

Besides these Zeeman-splitted resonances, several others show up
in the spectrum. We have identified with a +$(-)$ sign 
these involving changes $\Delta |L|$= 1($-1$) in the  
total orbital angular momentum\cite{note} with respect to that of
the gs.

At $B\sim$ 8 T, the positive $B$-dispersion brach of the
$\Delta n$= 0 resonance disappears, and a very low-lying,
positive
$B$-dispersion branch shows up. The origin of this transition
is the magnetic-induced change in the gs, as it can be
easily inferred looking at the $n$= 0 sp levels plotted in
Fig. \ref{fig2} and using the dipole selection rule to identify
the ones involved in the non spin-flip excitation. A
similar transition occurs at $B\sim$ 14 T. They are the
microscopic explanation of the appearance and disappearance
of the 'ridges' shown in Fig. \ref{fig3}, also found for
few electron nanorings\cite{Hal96,Emp99}.
It is worthwhile to notice that the rich structure appearing
in these nanorings (see below the $N$= 1 case) is
a peculiarity that has its origin in the smallness of $N$.
When $N$ is just a few tens, many electron-hole
pairs contribute to the building of the resonances and no
drastic changes appear in the FIR spectrum\cite{Emp99}.

We have also looked at the $N$= 1 ring for which some
experimental information is also available\cite{Lor99b}. As in
the $N$= 2 case, we have fixed $\omega_0$ so as to reproduce
the high energy resonance at $B$= 0. This yields $\omega_0\sim$
13.5 meV.

Figure \ref{fig5} shows several $n$= 0 sp levels of the $N$= 1 ring.
 For this system,
the total energy $E(1)$ is simply the energy of the lowest sp
level, $E(1)= \mu(1)$.
This has been used to calculate the addition
energy $\mu(2) = E(2) - E(1)$ shown in Fig. \ref{fig8}.
The dipole charge density strength and the energy of the more intense
$N$= 1 CDE's are plotted in Figs. \ref{fig6} and \ref{fig7}
as a function of $B$.

We thus see that the experimental data on FIR transmission spectroscopy
reflects that the surface ring morphology of the experimental
samples has indeed being translated to a true underlying electronic
ring structure\cite{Lor99}, and that a fair quantitative agreement
can be found between TDLSDFT calculations and experiment. Our 
calculations also give support to the way the experimental
resonances have been grouped, with the only doubt of the 'dot'
peak at $B$= 6 T and $\omega \sim$ 16.1 meV which could also be
a 'triangle' peak of $(-)$ character because in this region both
branches merge.
To unambiguously arrange the  peaks into branches and
disentangle the $B$ dispersion of the  modes, it would
be essential to experimentally   assign the polarization 
state to the main CDE's, as is has been done for antidot
arrays\cite{Bol95}.  This
is crucial in the analysis of the theoretical FIR response,
which otherwise does not allow us to distinguish between
peak fragmentation and different plasmon branches in some cases.

From the theory viewpoint, the main shortcomings are the 'cross' peaks
appearing at around 8 T for $N$= 2, and 10 T for $N$= 1, as well as
clear overestimation of the peak energy of the $(-)$ high energy 
$\Delta n$= 1 mode, which also lacks of some strength. 
These drawbacks are also qualitatively present in the
calculations of Ref. \onlinecite{Hal96}. It is alike that using
other possible confining potentials, like a jellium ring\cite{Emp99}
or that of Ref. \onlinecite{Tan99} which yields analytical
sp wave functions in the non-interacting case will improve the agreement
in view of other possible sources of uncertainty, as for example the
precise value of the ring radius $R_0$ (we have taken that of Ref.
\onlinecite{Lor99}, but larger values would also be acceptable), 
and the values of $m^*$, $g^*$ and $\epsilon$ corresponding to InAs.
In particular, the effective mass value seems to depend on whether
it is extracted from capacitance of from FIR spectroscopy.
We have checked that if we take\cite{Fri96,Mil97}
$m^*$= 0.08 we achieve a
better description of the 'dot' peaks in Figs. \ref{fig3} and
\ref{fig6} at the price of spoiling the description of 'diamond' and
'triangle' peaks. Yet, the patterns look qualitatively similar.

Finally, we show in Fig. \ref{fig8} the addition energies of both
rings as compared with the gate voltage shift of the lowest
capacitance maximum\cite{Lor99,Lor99b}. It can be seen that the
agreement between theory and experiment is rather poor. At 
$B \sim $ 12 T the calculations underestimate the shift voltage
around a factor of 3 for $N$= 2, and of 2 for $N$= 1.
We recall that the agreement
between capacitance spectroscopy experiments and exact diagonalization
calculations of few electron quantum dots is also only qualitative
\cite{Ash93,Pal94}. We cannot discard that using a different radius 
$R_0$ for each ring would not improve the agreement but have not
tried this possibility to avoid  too much parameter fitting 
in the calculation.
The electron-electron interaction determines the energy difference
between $\mu(1)$ and $\mu(2)$ at $B$= 0.
A small bump in $\mu(2)$ at $B\sim$ 2-3 T is the signature of full
polarization. A similar structure shows up in the experimental
points but between 3-4 T. Interestingly however, the change
in the electronic structure at $B\sim$ 8 T is visible in the calculated
addition energy $\mu(2)$.

This work has been performed under grants PB95-1249 from CICYT,
Spain, and 1998SGR00011 from Generalitat of Catalunya.
A. E. acknowledges support from the DGES (Spain), and A. L. 
from the German Ministry of Science (BMBF).

\begin{figure}
\caption{
Total energy of the $N$= 2 ring as a function of $B$. The dashed
line corresponds to an $S_z$= 0 gs, and the solid line to an 
$S_z$= 1 gs. The ring becomes fully polarized near $B$= 3 T.
}
\label{fig1}
\end{figure}
\begin{figure}
\caption{Several $n$= 0 sp energies for the $N$= 2 ring as a
function  of $B$. The quantum labels ($n, l, \sigma$)
of the sp states  are also  indicated.
}
\label{fig2}
\end{figure}
\begin{figure}
\caption[]{Charge density strength function 
vs excitation energy (solid lines)
for $N$= 2 and $B$= 0 to 15 T. The symbols represent the 
experimental peak energies\cite{Lor99}. The dashed
line at $B$= 0, 1 and 2 T is the longitudinal spin density
strength function.  
}
\label{fig3}
\end{figure}
\begin{figure}
\caption[]{Energy of the more intense CDE's as a function of $B$
for $N$= 2. The dashed line represents the cyclotron frequency 
$\omega_c$, and the solid
lines are drawn to guide the eye. The thick symbols represent
the experimental data\cite{Lor99}.
}
\label{fig4}
\end{figure}
\begin{figure}
\caption{Several $n$= 0 sp energies for the $N$= 1 ring as a
function  of $B$. The lower energy state of
each $(n,l)$ pair has spin up. 
}
\label{fig5}
\end{figure}
\begin{figure}
\caption[]{Same as Fig. \ref{fig3} for $N$= 1. The experimental
data are from Ref. \onlinecite{Lor99b}
}
\label{fig6}
\end{figure}
\begin{figure}
\caption[]{Same as Fig. \ref{fig4} for $N$= 1. The experimental
data are from Ref. \onlinecite{Lor99b}
}
\label{fig7}
\end{figure}
\begin{figure}
\caption[]{Addition energies as a function of $B$ (left vertical scale).
The symbols are the experimental capacitance 
data\cite{Lor99,Lor99b} (right vertical scale).
Large dots correspond to $N$= 2, and small dots to $N$= 1.
}
\label{fig8}
\end{figure}

\begin{references}
\bibitem{note} We have adopted the convention that $l$ represents
the sp orbital angular momentum changed of sign \cite{Fer94}. The
applied $B$ points towards the positive direction of the $z$ axis.
Consequently, most occupied sp levels have a negative orbital
angular momentum, the total orbital angular momentum $L_z$
is  zero or negative, and the total spin $S_z$ is zero or positive.
\end{references}
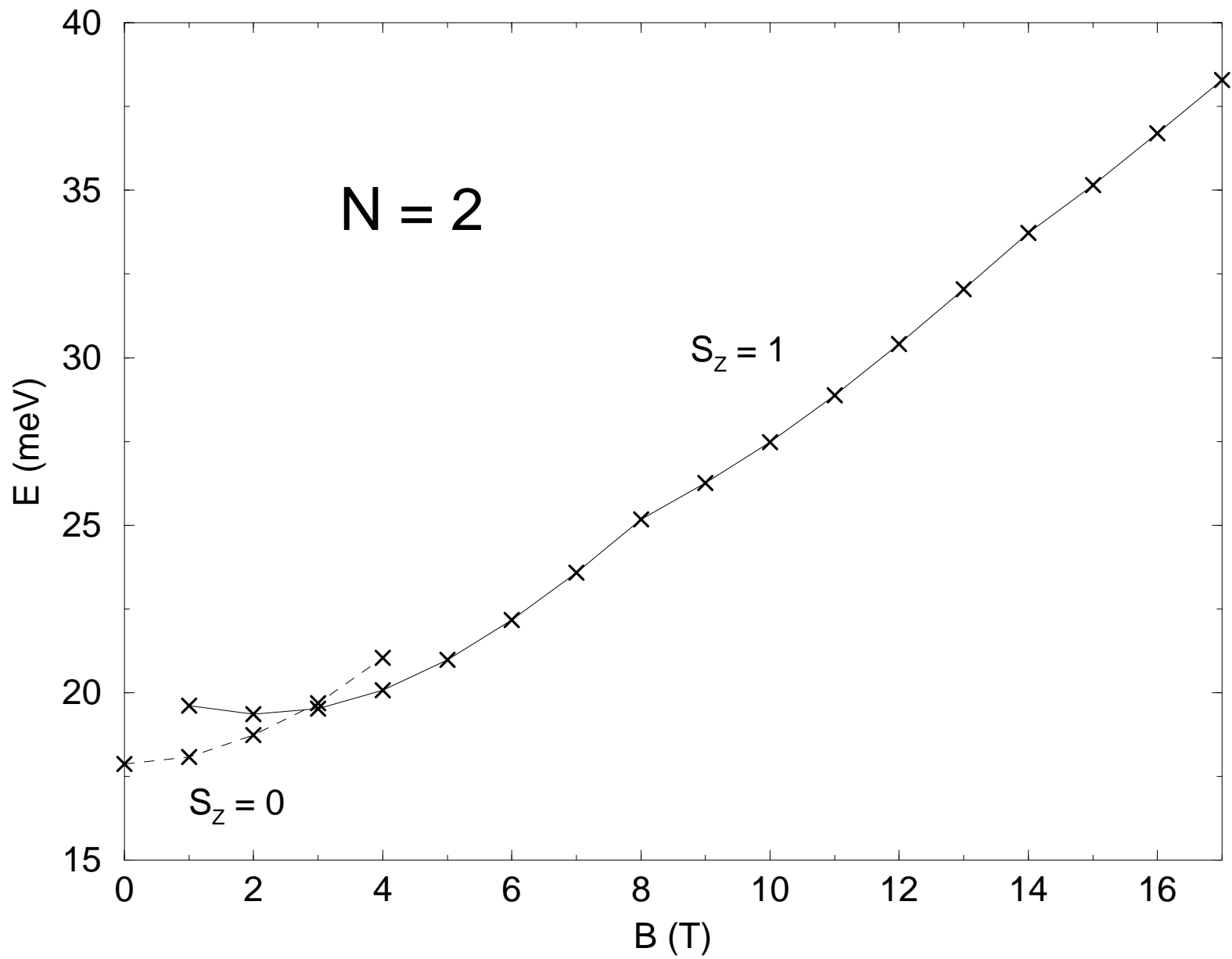
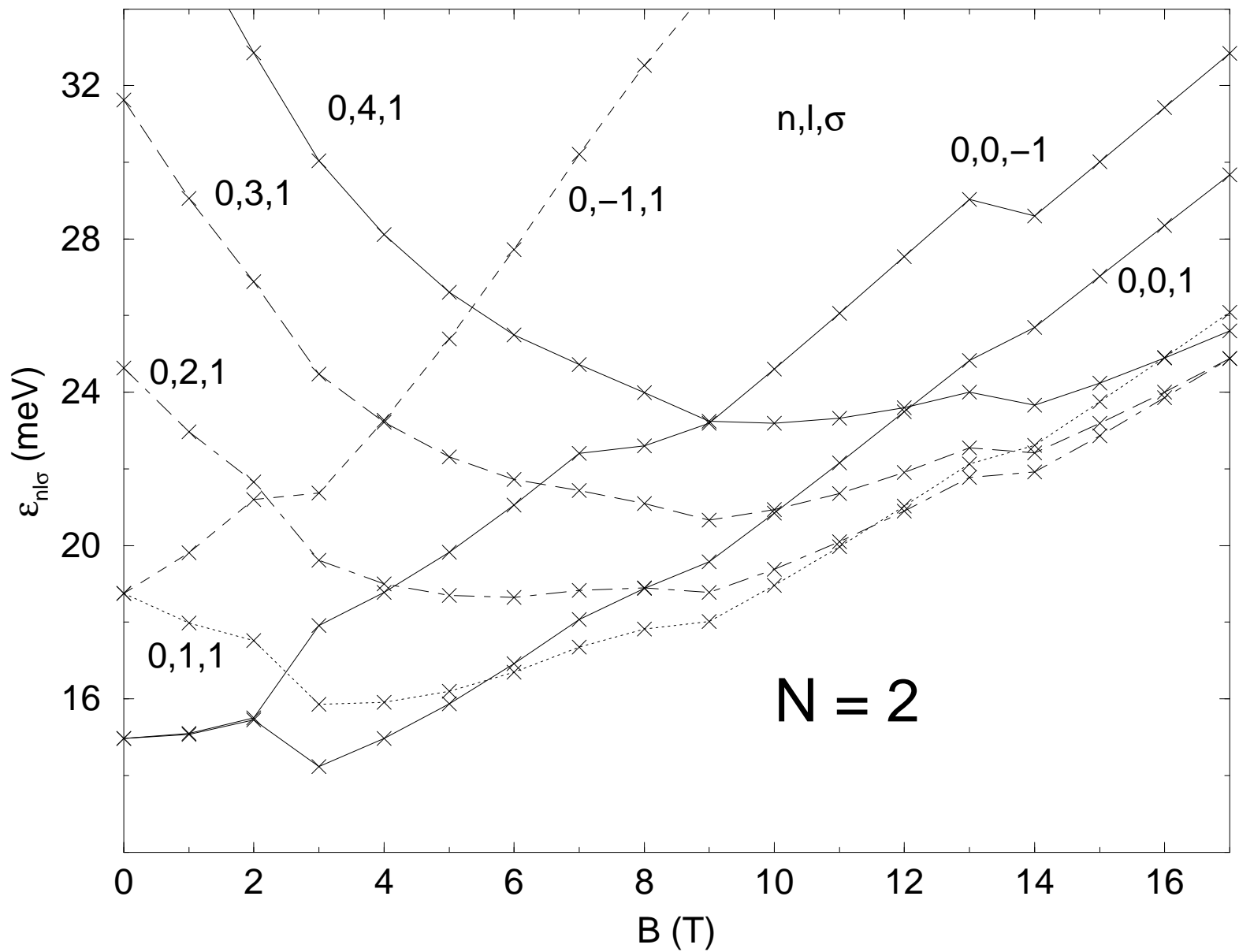
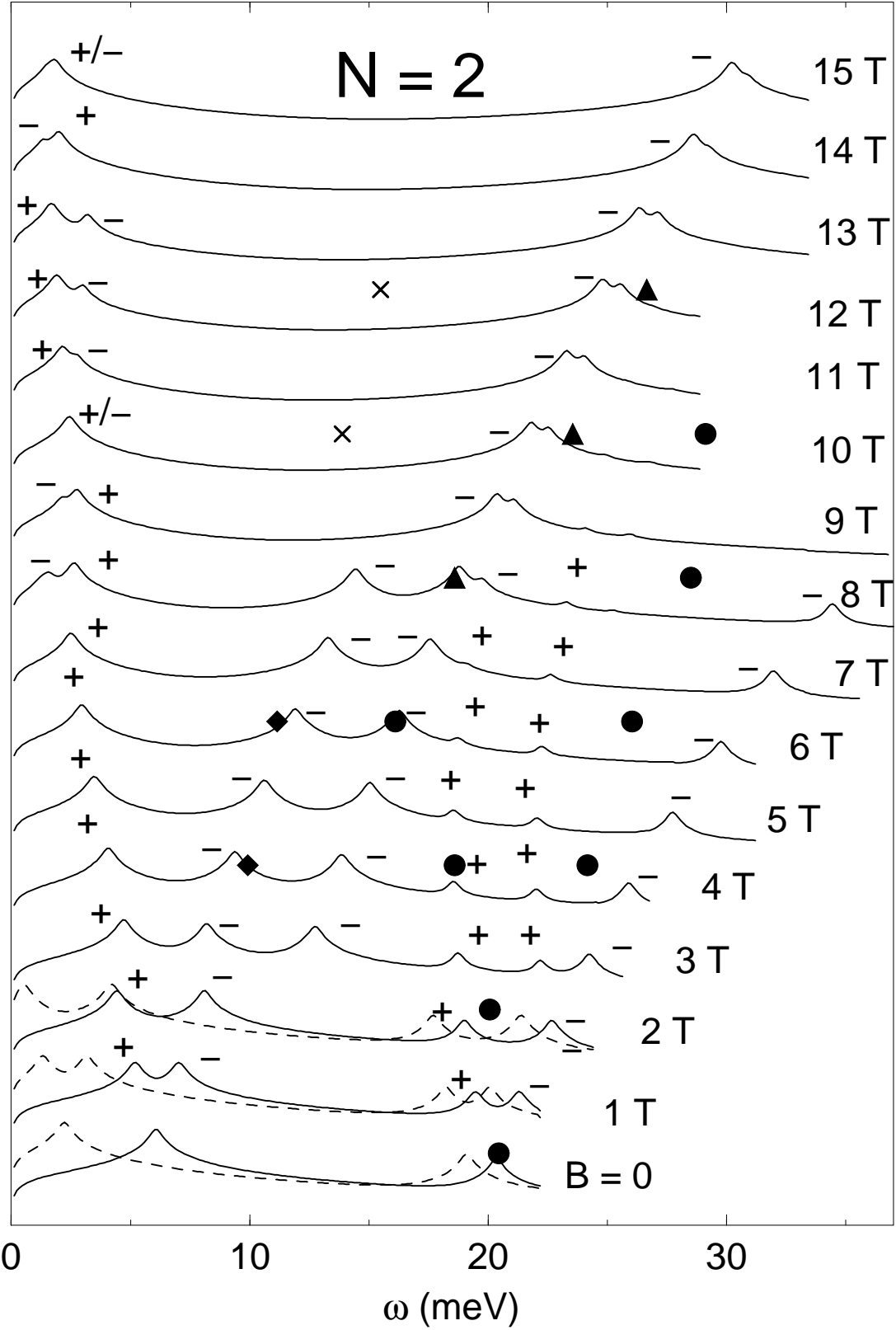
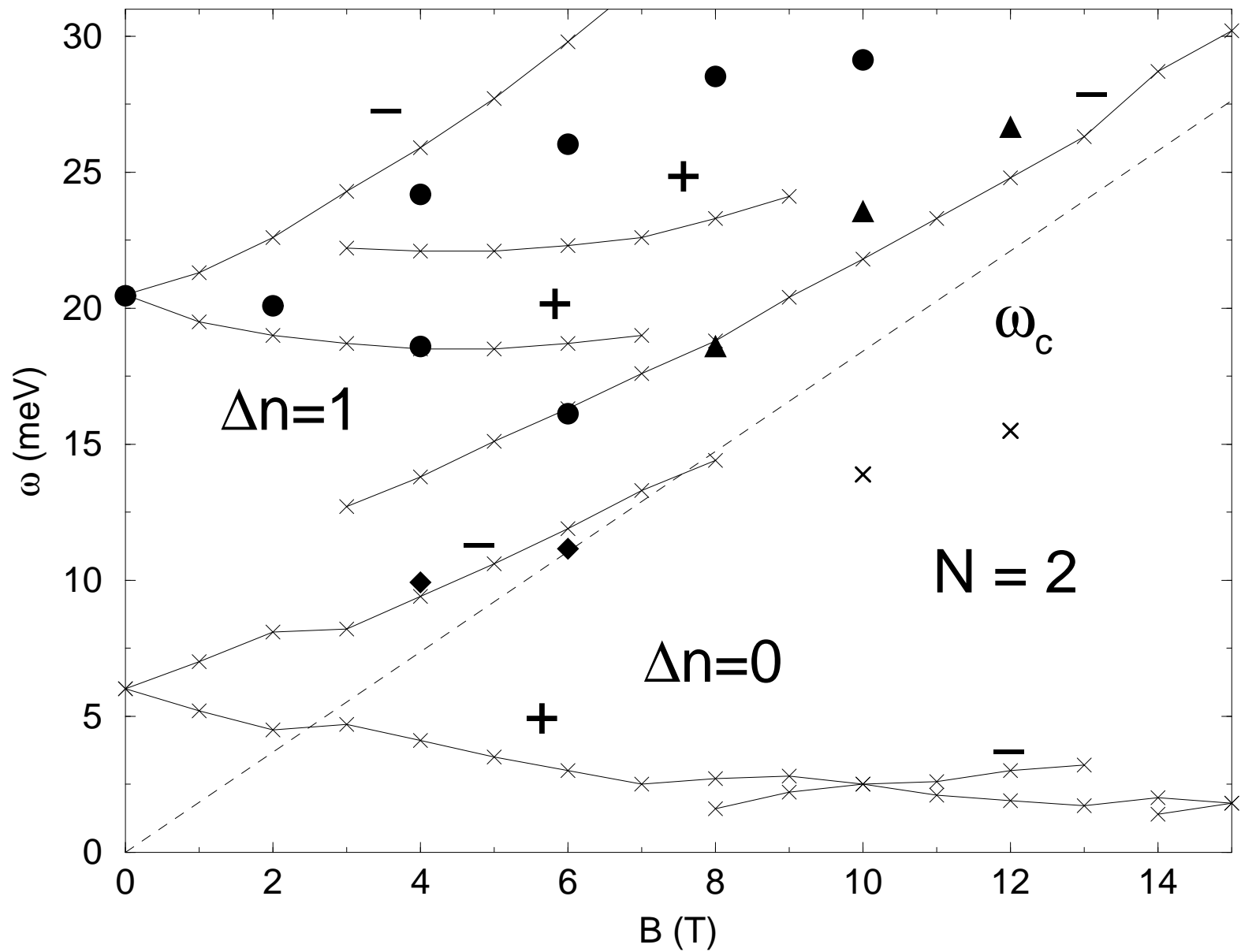
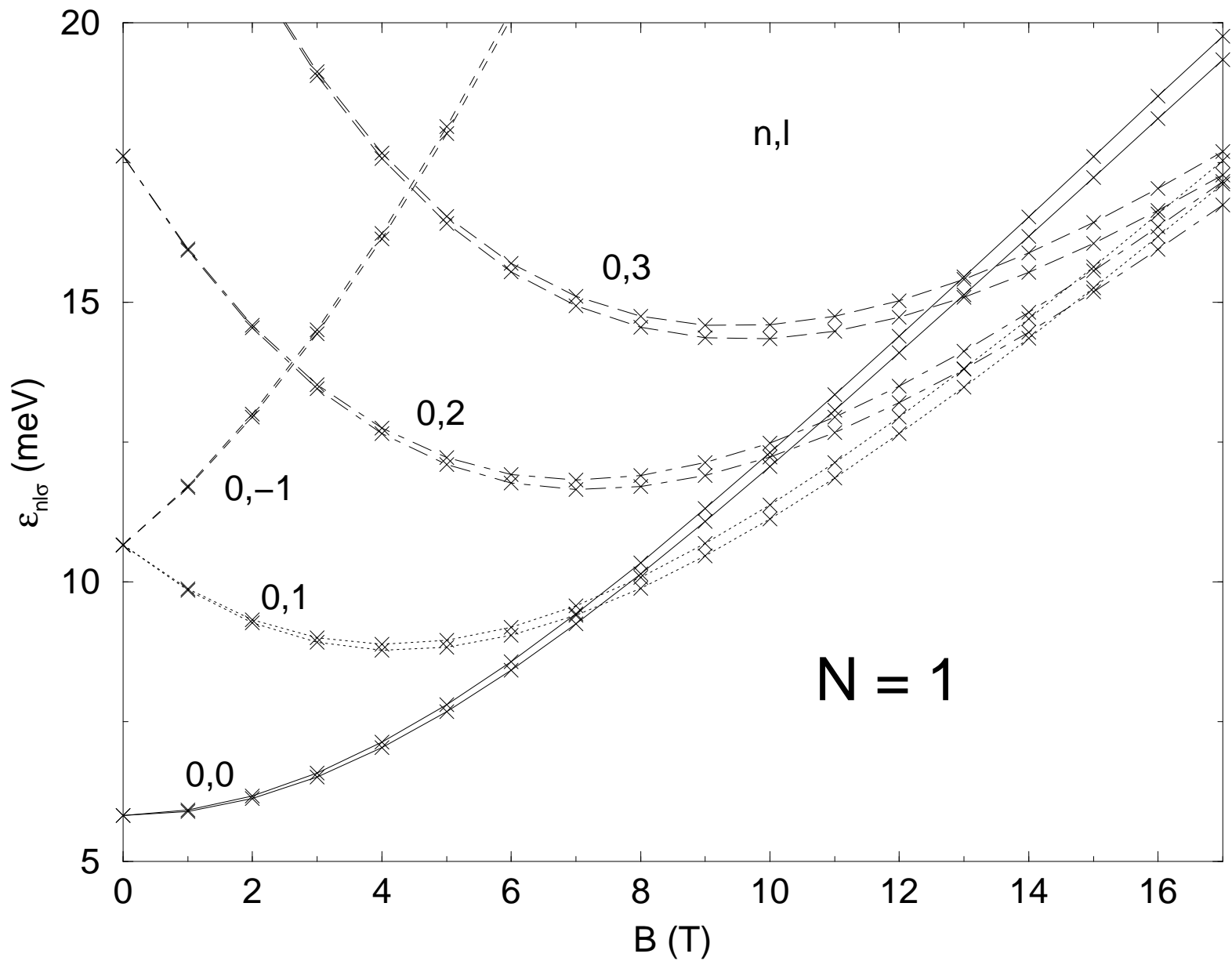
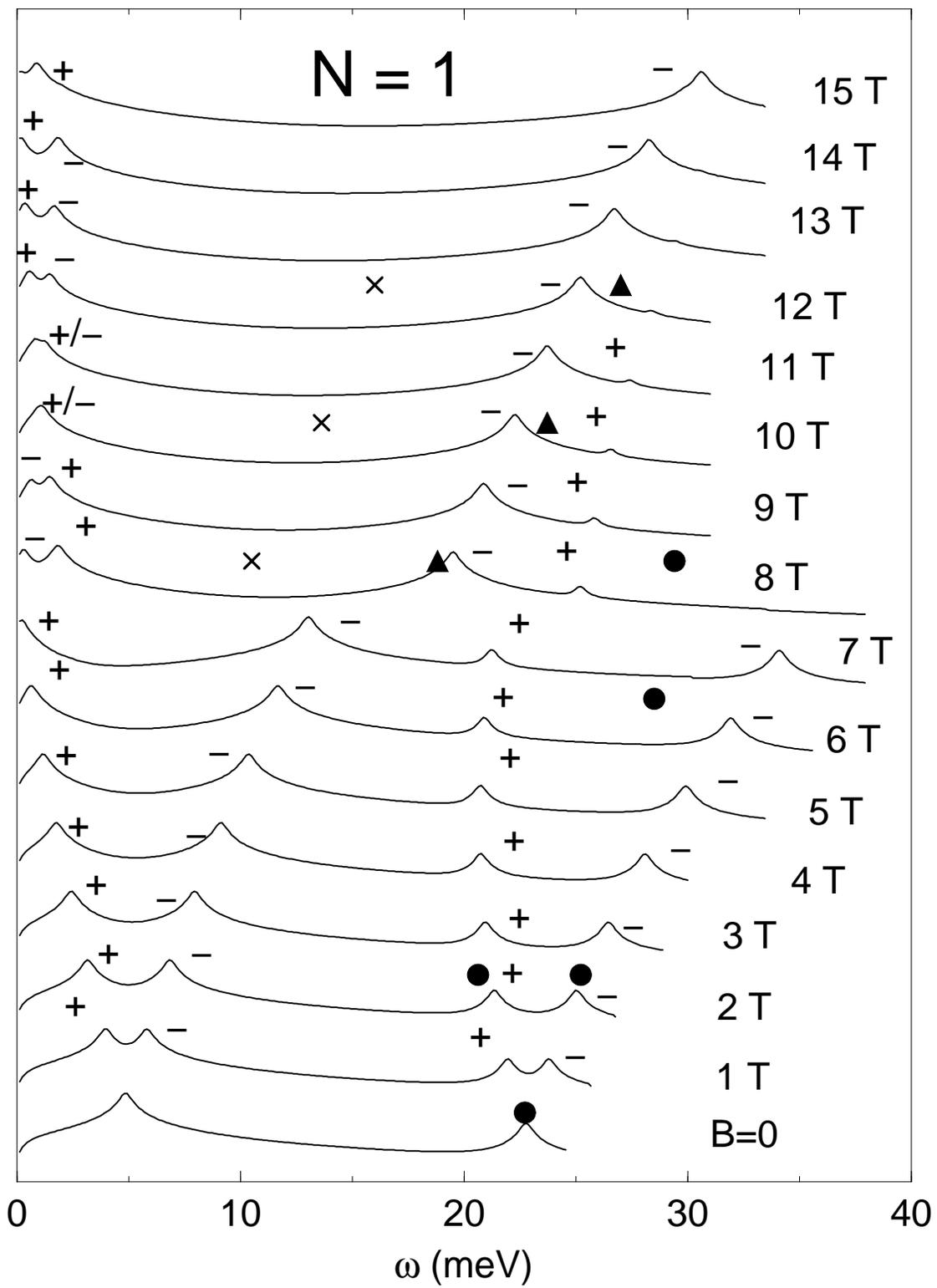
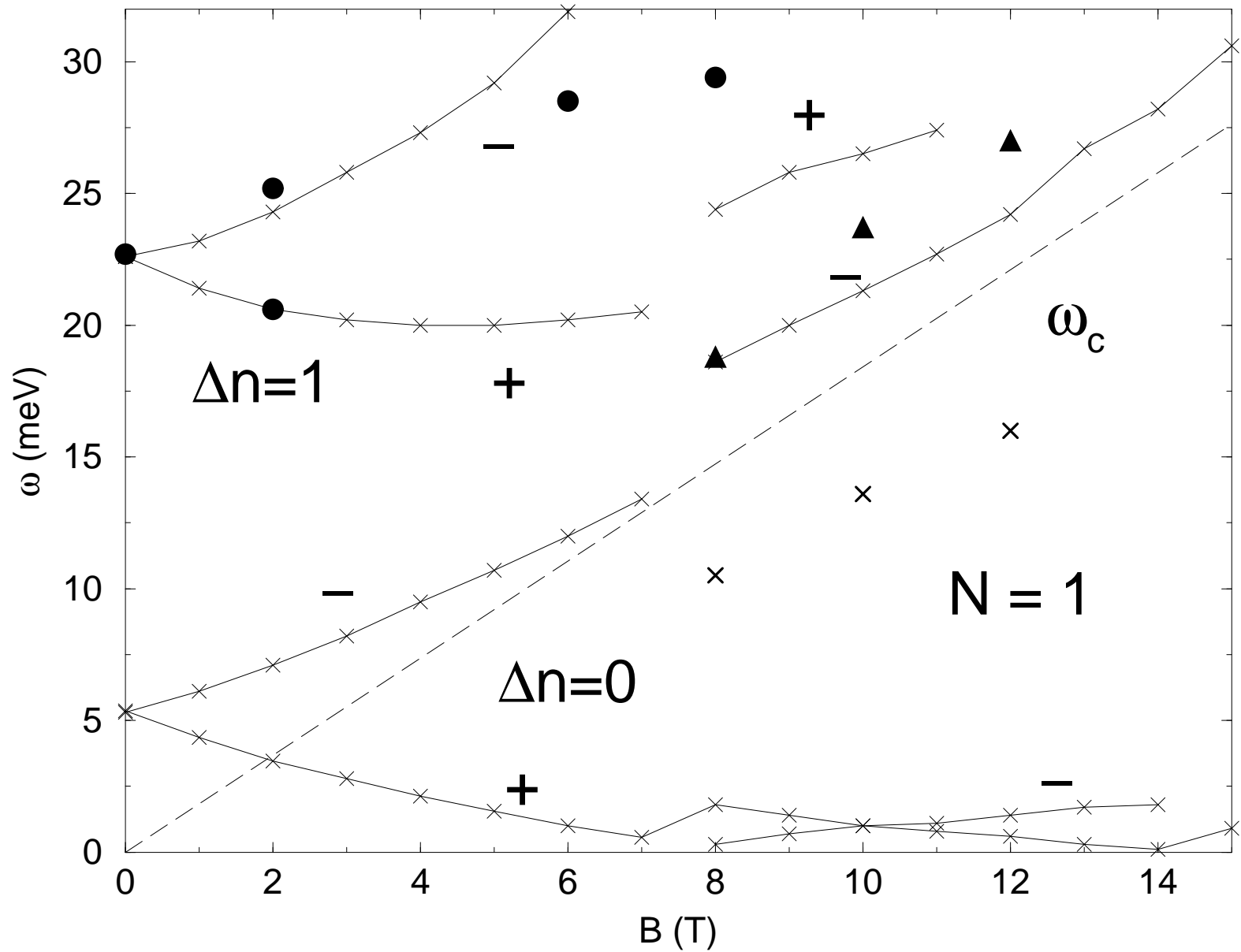
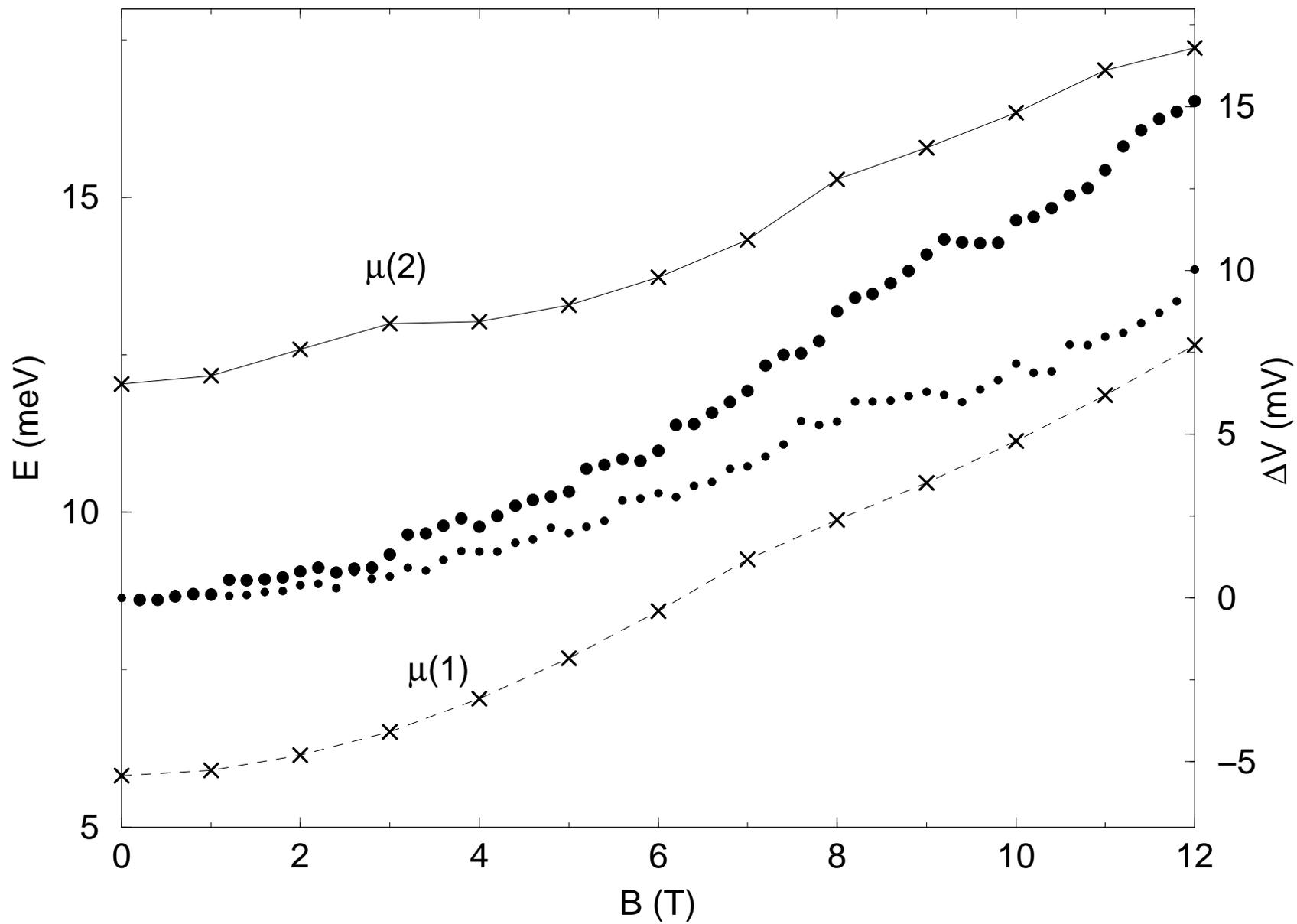
\end{document}